%% file: main.tex
\documentclass[10pt,conference]{IEEEtran}
\IEEEoverridecommandlockouts

\usepackage{cite}
\usepackage{amsmath,amssymb,amsfonts}
\usepackage{amsthm}
\usepackage{algorithmic}
\usepackage{graphicx}
\usepackage{textcomp}
\usepackage{xcolor}
\def\BibTeX{{\rm B\kern-.05em{\sc i\kern-.025em b}\kern-.08em
    T\kern-.1667em\lower.7ex\hbox{E}\kern-.125emX}}

\usepackage[caption=false,font=footnotesize]{subfig}
\usepackage{enumitem}
\usepackage{booktabs}
\usepackage{tikz}
\usepackage{listings}
\usepackage{xspace}
\usepackage{framed}
\usepackage{float}

\newcommand\algorithmicprocedure{\textbf{procedure}}
\newcommand{\algorithmicendprocedure}{\algorithmicend\ \algorithmicprocedure}
\makeatletter
\newcommand\PROCEDURE[3][default]{%
  \ALC@it
  \algorithmicprocedure\ \texttt{#2}(#3)%
  \ALC@com{#1}%
  \begin{ALC@prc}%
}
\newcommand\ENDPROCEDURE{%
  \end{ALC@prc}%
  \ifthenelse{\boolean{ALC@noend}}{}{%
    \ALC@it\algorithmicendprocedure
  }%
}
\newenvironment{ALC@prc}{\begin{ALC@g}}{\end{ALC@g}}
\makeatother

\newenvironment{framedsized}[1][\hsize]
  {\MakeFramed{\hsize#1\advance\hsize-\width \FrameRestore}}%
  {\endMakeFramed}

\tikzset{place/.style={circle,draw,minimum size=1.5em}}
\tikzset{node distance=5em}

\newtheorem{definition}{Definition}

\newcommand{\techname}{OrdinalFix\xspace}

\usepackage[hidelinks]{hyperref}

\begin{document}

\title{\techname: Fixing Compilation Errors via Shortest-Path CFL Reachability\\{\footnotesize \thanks{{*}Corresponding author.}}}

\author{
  \IEEEauthorblockN{
    Wenjie Zhang\IEEEauthorrefmark{2},
    Guancheng Wang\IEEEauthorrefmark{2},
    Junjie Chen\IEEEauthorrefmark{3},
    Yingfei Xiong\IEEEauthorrefmark{2},
    Yong Liu\IEEEauthorrefmark{4},
    Lu Zhang\IEEEauthorrefmark{2}*}
  \IEEEauthorblockA{
    \IEEEauthorrefmark{2}
    Key Laboratory of High Confidence Software Technologies (Peking University), MoE\\
    School of Computer Science, Peking University, P. R. China\\
    \IEEEauthorrefmark{3}College of Intelligence and Computing, Tianjin University, P. R. China\\
    \IEEEauthorrefmark{4}College of Information Science and Technology, Beijing University of Chemical Technology, P. R. China\\
    zhang\_wen\_jie@pku.edu.cn, guancheng.wang@pku.edu.cn, junjiechen@tju.edu.cn,\\
    xiongyf@pku.edu.cn, lyong@mail.buct.edu.cn, zhanglucs@pku.edu.cn
  }}




\maketitle

\begin{abstract}
  The development of correct and efficient software can be hindered by compilation errors, which must be fixed to ensure the code's syntactic correctness and program language constraints. Neural network-based approaches have been used to tackle this problem, but they lack guarantees of output correctness and can require an unlimited number of modifications. Fixing compilation errors within a given number of modifications is a challenging task. We demonstrate that finding the minimum number of modifications to fix a compilation error is NP-hard. To address compilation error fixing problem, we propose OrdinalFix, a complete algorithm based on shortest-path CFL (context-free language) reachability with attribute checking that is guaranteed to output a program with the minimum number of modifications required. Specifically, OrdinalFix searches possible fixes from the smallest to the largest number of modifications. By incorporating merged attribute checking to enhance efficiency, the time complexity of OrdinalFix is acceptable for application. We evaluate OrdinalFix on two datasets and demonstrate its ability to fix compilation errors within reasonable time limit. Comparing with existing approaches, OrdinalFix achieves a success rate of 83.5\%, surpassing all existing approaches (71.7\%).
\end{abstract}

\begin{IEEEkeywords}
  Compilation Error, CFL Reachability
\end{IEEEkeywords}

\input{chaps/01-intro}

\input{chaps/02-problem-formulation.tex}

\input{chaps/03-np-completeness.tex}

\input{chaps/04-approach.tex}

\input{chaps/06-experiments.tex}

\input{chaps/07-related-work.tex}

\input{chaps/08-conclusion.tex}

\section*{Acknowledgment}

The work is supported in part by the National Key Research and Development Program of China under Grant 2022YFB4501902 and the National Natural Science Foundation of China under Grant Nos. 62232003, 62232001, 61902015, and 62002256.

\bibliographystyle{IEEEtran}
\bibliography{IEEEabrv,main}

\clearpage
\input{chaps/09-algorithm.tex}

\end{document}

%% file: chaps/01-intro.tex
\section{Introduction}

Errors are inevitable in software development. Manual error fixing is a time-consuming and error-prone process, especially when software becomes larger and more complex. Therefore, automated error fixing has attracted extensive attention in recent years~\cite{DBLP:conf/icse/GazzolaMM18}. In the field of automated error fixing, most research focuses on fixing logic errors in compilable programs and ignores a large class of errors occurring at the stage of compiling, i.e., compilation errors. Compilation errors are frequent in practice and typically due to inexperience of programmers or lack of attention to details, such as missing scope delimiters~\cite{DBLP:conf/aaai/GuptaPKS17}. According to an existing study at Google~\cite{DBLP:conf/icse/SeoSEAB14}, both novices and experienced developers make such errors. Also, more and more tools are designed for processing code snippets from the web, where compilation errors are common. Therefore, automated compilation-error fixing is a critical task.

For a typical programming language, compilable programs need to satisfy two types of constraints: syntactic constraints and semantic constraints. The syntactic constraints ensure that the given program conforms to some context-free grammar (CFG). The semantic constraints require that the program passes some checks for non-context-free properties, such as definition-and-use relations of variables and type matching relations in expressions. We should mention that, in this paper, our use of semantic constraints is focused on representing compiler constraints, rather than ensuring the program's functional correctness.

Given a non-compilable program, which violates some syntactic constraints or semantic constraints, or both, the task of fixing compilation errors in the program is to modify the program to change it into a compilable form. Typically, it is not expected that we massively change the program since such a way of modifying the program would have the changed program deviate largely from the original program. In an extreme case, if we delete all the tokens in the non-compilable program, what we obtain is a compilable but useless program. Therefore, in this paper, we define the problem of fixing compilation errors in a program as the problem of finding a minimum number of changes to make the program compilable.

Over the years, several approaches have been proposed for helping solve this problem. Parsing-stage-based approaches (see e.g.,~\cite{DBLP:journals/toplas/CorchueloPRT02,DBLP:conf/pldi/ParrF11}) provide fixing recommendations by analyzing the surrounding code of the error reported in the error message. Some of these approaches have been integrated into mature IDEs such as Eclipse. However, the effectiveness of these approaches is limited due to inaccurate error messages. Machine-learning-based approaches (see e.g.,~\cite{DBLP:journals/corr/BhatiaS16,DBLP:conf/sigsoft/0001RJGA19}) predict fixes for compilation errors via learning from known fixes using some machine learning algorithms. However, the capability of these approaches highly depends on the sufficiency of the training dataset and thus the effectiveness is also limited. Furthermore, these approaches cannot guarantee fixing the errors with minimum changes. In the field of algorithms, given a context-free grammar (denoted as $CF$) and a string (denoted as $p$) such that $p \notin L(CF)$, Aho and Peterson\cite{DBLP:journals/siamcomp/AhoP72} proposed an algorithm for calculating a modified string (denoted as $p'$) with minimum edge distance, and demonstrated that the asymptotic execution time of the algorithm is $O(n^3)$, where $n$ is the length of $p$ (i.e., $|p|$). In principle, when we are facing only violations of syntactic constraints, this algorithm is applicable and guarantees to find fixes with minimum changes. However, this algorithm is not applicable when there are violations of semantic constraints. Furthermore, even if there are only syntactic errors, this algorithm is not able to guarantee that the resulting program satisfies all required semantic constraints.

In fact, fixing semantic errors is more complex than fixing syntactic errors. Fixing compilation errors for common programming languages, such as Java, is NP-hard, as shown in Section~\ref{sec:NPproof}. That is to say, to fix compilation errors, there is no complete algorithm that guarantees to produce minimum modifications in polynomial time unless P=NP. A na\"ive algorithm for compilation error fixing is to enumerate all syntactically correct fixes from fewer modifications to more modifications with a syntactic error fixing algorithm (e.g., \cite{DBLP:journals/siamcomp/AhoP72}) and then check the compilability with a compiler.
However, an exponential number of syntactically correct but semantically incorrect programs would be generated during fixing by the algorithm.

In real-world scenarios, addressing compilation errors through human intervention commonly takes a few seconds to several minutes~\cite{DBLP:conf/icse/SeoSEAB14}. Consequently, any algorithm designed to assist developers in compile error fixing should operate within a comparably brief time frame. Unfortunately, the previously mentioned na\"ive algorithm does not fulfill this requirement due to the rapid increase in potentially valid but semantically erroneous solutions, which leads to an exponential growth.

In order to reduce the time consumption of compilation error fixing, we propose \techname\footnote{The source code and data of \techname is available at \url{https://github.com/myxxxsquared/OrdinalFix}}, which tries to perform a more fine-grained semantic checking with merging and branching reduction, so that the algorithm can complete in an acceptable period of time.

For syntactic analysis, modern compilers typically rely on LL-based or LR-based parsers to efficiently analyze program structures. These parsers can accurately process correct programs. However, for erroneous programs, they only contain a single parse state when encountering errors, which may lead to ignoring possible modifications that could fix the error. Although there are approaches to recovering from errors during parsing, they do not consider all possible fixes. To explore all possible fixes, we utilize a modified shortest-path CFL reachability algorithm. Although this algorithm has a higher time complexity than LL-based or LR-based parsers, it can capture all possible fixes.

For semantic constraints, we utilize the attribute grammar approach commonly used in compiler construction to perform attribute checking in CFL reachability. We encode semantic constraints into attributes and perform merged attribute checking on top of each reduction relationship obtained via CFL reachability on the modification graph. This approach allows us to effectively handle semantic constraints and ensure that the modified program satisfies all constraints.

We evaluate \techname on fixing compilation errors with two datasets, consisting of a synthesized dataset in an object-oriented language (i.e., Middleweight Java) and a real world dataset in a procedural language (i.e., C). The empirical results show that, for the Middleweight Java programs, \techname is able to fix all the programs within 600 seconds. For C programs the median time for fixing a function body is only 0.6 seconds, demonstrating the efficiency of \techname. Furthermore, we compare the fixing rate of \techname with that of existing approaches on the same C dataset, where \techname achieves a success rate of 83.5\% (5,757 out of 6,978) for programs with compilation errors within 600 seconds, surpassing all existing approaches.


To sum up, we make the following contributions in this paper:
\begin{itemize}
    \item We demonstrate the NP-hardness of the problem of fixing compilation errors with the minimum number of modifications.
    \item We propose a fine-grained algorithm (named \techname) for fixing compilation errors based on attribute checking over the CFL reachability graph, where we rely on merging and pruning to reduce its time consumption.
    \item We implement \techname for compilation error fixing and evaluate \techname on generated programs written in Middleweight Java and real-world C programs with compilation errors, demonstrating the effectiveness and efficiency of \techname.
\end{itemize}

%% file: chaps/02-problem-formulation.tex
\section{Problem Formulation}\label{sec:problemformulation}

When fixing compilation errors, it is typical that we want to modify the erroneous program as little as possible.
That is, given a programming language, which is defined by some context-free grammar and some semantic constraints in form of typing rules and a program $p$ that does not satisfy the syntactic constraints and/or the semantic constraints, the goal is to find a program $p'$ that is closest to $p$ to make $p' \in L(CF)$ established and $p'$ satisfies the semantic constraints. It means that we are able to get $p'$ from $p$ by performing some minimum modifications on $p$.

\begin{definition}[compilation error fixing]\label{def:problem}
    Given a programming language and a program $p$ such that $p$ does not compile, the problem of fixing compilation errors in $p$ is to find program $p'$ such that $p'$ compiles and the edit distance between $p$ and $p'$ is minimum, i.e., for all $p''$ that compiles, $d(p,p') \leq d(p,p'')$, where $d(p_1,p_2)$ is the edit distance between $p_1$ and $p_2$.
\end{definition}

Here, the edit distance between two programs $p_1$ and $p_2$ accounts for the number of modifications required to transform $p_1$ into $p_2$. These modifications encompass token insertion, deletion, or replacement operations.

For the ease of presentation, we assume that the empty program always compiles. This assumption is reasonable because the empty program means a program doing nothing. Under the assumption, for any program $p$ that does not compile, we can always find a fix with $|p|$ removal operations, where $|p|$ is the number of tokens in $p$. Usually, the empty program is not the fix we want, because it means to remove the whole program. However, the empty program gives an upper bound of compilation error fixing. Any program $p$ can be fixed within $|p|$ modifications.

We also assume that the context-free grammar of a programming language is unambiguous, thus a program $p$ will have at most one parse tree.

%% file: chaps/03-np-completeness.tex
\section{NP Hardness}

\label{sec:NPproof}

According to an earlier study, finding fixes that conform to context-free grammars with the minimum number of modifications can be done within $\mathrm O(n^3)$~\cite{DBLP:journals/siamcomp/AhoP72}; however, it would be much more complex to find fixes that are compilable. Below, we show that for a simple object-oriented language that allows local variable declarations, finding fixes that satisfy the typing semantic constraints with the minimum number of modifications is NP-hard. Specifically, we reduce the maximum independent set problem to the compilation error fixing problem.

The maximum independent set (MIS) is defined as follows. Given an undirected graph $G=(V, E)$, the problem is to find the largest subset of vertices $V' \subset V$ such that any two vertices in $V'$ are not adjacent in the graph $G$. MIS is proven to be NP-complete.

Given a MIS problem, let us denote the graph in the MIS problem as $G=(V, E)$. We reduce the MIS problem to a problem of fixing compilation errors in a Java-like language. \figurename~\ref{fig:mjdefs} depicts the external declaration of two classes (i.e., \textit{InMIS} and \textit{OutMIS}). This declaration ensures that \textit{a.addEdge(b)} is legal based on the typing semantic constraints only if either \textit{a} or \textit{b} is of type \textit{OutMIS}.

\begin{figure}[t]
    \subfloat[External declarations\label{fig:mjdefs}]{
        \begin{minipage}{.58\linewidth}
\begin{lstlisting}[language=Java,basicstyle=\ttfamily\fontsize{8}{10}\selectfont,frame=single]
class InMIS {
    void addEdge(OutMIS obj);
}
class OutMIS extends InMIS {
    void addEdge(InMIS obj);
}
\end{lstlisting}
        \end{minipage}
    }
    \hfil
    \subfloat[Program\label{fig:mjcode}]{
        \begin{minipage}{.8\linewidth}
\begin{lstlisting}[language=Java,basicstyle=\ttfamily\fontsize{8}{10}\selectfont,frame=single,linewidth=.4\linewidth]
InMIS v1;
InMIS v2;
InMIS v3;
...
v2.addEdge(v3);
v2.addEdge(v3);
...
\end{lstlisting}
        \end{minipage}
    }
    \caption{Program to fix for MIS.}
\end{figure}

Based on this declaration, we construct a program snippet $p$ with $(n+mn)$ lines (depicted in \figurename~\ref{fig:mjcode}), where $n$ is the number of vertices in $G$, and $m$ is the number of edges in $G$. Specifically, program snippet $p$ consists of two parts, which have $n$ and $mn$ lines, respectively. For the first $n$ lines, the content of the $i$-th line is ``\textit{InMIS v\{i\};}'', representing the $i$-th vertex in $G$. Here, \textit{\{i\}} means replacing with the number of $i$; for example, the third line is ``\textit{InMIS v3;}''. The following $mn$ lines range from line $(n+1)$ to line $(n+mn)$. Given $1\leq k\leq m$, let the $k$-th edge in $G$ be in the form of $(v_x, v_y)$, from line $(kn+1)$ to line $(kn+n)$, the content is the same, i.e., ``\textit{v\{x\}.addEdge(v\{y\});}''. For example, if the first edge in the graph is $\left(v_2, v_3\right)$, from line $(n+1)$ to line $2n$, the content of each line is ``\textit{v2.addEdge(v3);}''.

In $p$, the first $n$ lines define $n$ variables for the $n$ vertex in $G$. The remaining $mn$ lines represent the edge constraints. It is obvious that $p$ does not compile as long as the edge set of the graph is not empty. Each line after the first $n$ lines has a typing error, because \textit{InMIS.addEdge(InMIS)} is not defined in the external declaration. If we modify some of the first $n$ lines by changing \textit{InMIS} to \textit{OutMIS}, the program may compile, because \textit{InMIS.addEdge(OutMIS)}, \textit{OutMIS.addEdge(InMIS)}, and \textit{OutMIS.addEdge(OutMIS)} are all defined in the declaration.

Based on the preceding analysis, if the set of variables changed by a fix of $p$ with a minimum number of modifications is $V'$, $V-V'$ must be a maximum independent set; otherwise, a larger independent set would lead to a fix of $p$ containing fewer modifications. Similarly, a maximum independent set in $G$ also corresponds to a fix with a minimum number of modifications. Therefore, the maximum independent set problem is reducible to the compilation error fixing problem. That is to say, the problem of finding a minimum fix that compiles is NP-hard.

If we have a closer look at the program reduced from the maximum independent set problem, we know that the NP-hardness comes from the fact that there are up to an exponential number of different ways of variable declaration by modifying at most a polynomial number of places. In principle, type compliance should be checked against each way of variable declaration, incurring an exponential time complexity if $P\neq NP$.

%% file: chaps/04-approach.tex
\section{Preliminaries}

In this section, we review two existing techniques that \techname is built on: CFL reachability and attribute grammar.

CFL reachability is a parsing algorithm that captures all possible fixes for a given erroneous program. It has a higher time complexity than LL-based or LR-based parsers, but it can account for all potential modifications needed to fix the error.

Attribute grammar is a way to encode semantic constraints into attributes and perform attribute checking on top of each reduction relationship obtained via a parsing algorithm. This technique can help ensure that any modifications made during error fixing maintain the program's semantic correctness.

\subsection{CFL Reachability}

The CFL reachability algorithm is used to determine whether there exists a path between two given nodes in a directed graph, such that the symbol string of the path conforms to a context-free grammar. This algorithm was first proposed for interprocedural program analysis~\cite{DBLP:conf/slp/Reps97}.

To apply the CFL reachability algorithm, the grammar is first normalized into a normal form where the right-hand side of each production rule contains at most two symbols. This means that all production rules have one of the following three forms: $A \to \epsilon$, $A \to B$, or $A \to BC$, where $A$ represents a non-terminal symbol and $B$ or $C$ represents a terminal or a non-terminal symbol.

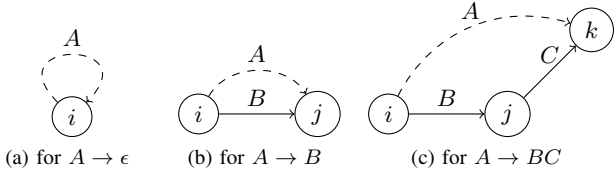
\begin{figure}[t]
    \centering
    \subfloat[for $A \to \epsilon$\label{fig:gen-0}]{
        \small
        \begin{tikzpicture}
            \node[place] (0) {$i$};
            \path[->] (0) edge [loop above,in=45,out=135,looseness=8,dashed] node {$A$} (0);
        \end{tikzpicture}}
    \hfil
    \subfloat[for $A \to B$\label{fig:gen-1}]{
        \small
        \begin{tikzpicture}
            \node[place] (0) {$i$};
            \node[place,right of=0] (1) {$j$};
            \path[->] (0) edge [above] node {$B$} (1);
            \path[->] (0) edge [above,bend left=60,dashed] node {$A$} (1);
        \end{tikzpicture}}
    \hfil
    \subfloat[for $A \to BC$\label{fig:gen-2}]{
        \small
        \begin{tikzpicture}
            \node[place] (0) {$i$};
            \node[place,right of=0] (1) {$j$};
            \node[place,above right of=1] (2) {$k$};
            \path[->] (0) edge [above] node {$B$} (1);
            \path[->] (1) edge [above] node {$C$} (2);
            \path[->] (0) edge [above,bend left=40,dashed] node {$A$} (2);
        \end{tikzpicture}}
    \caption{Edges caused by productions in CFL reachability.}
\end{figure}

Next, the algorithm iteratively adds new edges labeled with non-terminal symbols in the context-free grammar to the directed graph $G$, with three principles. 1) If a production rule is of the form $A \to \epsilon$, then self-loop edges labeled with $A$ are added to each vertex in the initialization stage, as shown in \figurename~\ref{fig:gen-0}. 2) If the production rule is of the form $A \to B$, a new edge labeled with $A$ is added from vertex $v_i$ to vertex $v_j$, where the original edge between $v_i$ and $v_j$ is labeled with $B$, as shown in \figurename~\ref{fig:gen-1}. 3) If the production rule is of the form $A \to BC$, a new edge labeled with $A$ is added from vertex $v_i$ to vertex $v_k$, where the original edges between $v_i$ and $v_j$ and between $v_j$ and $v_k$ are labeled with $B$ and $C$, respectively, as shown in \figurename~\ref{fig:gen-2}.

The algorithm maintains a work list to record changes and determine where to add edges caused by productions $A \to B$ and $A \to BC$, and it terminates when no more edges can be added.

Finally, the CFL reachability algorithm checks if there exists an edge labeled with the start symbol after the loop exits.

\subsection{Shortest-Path CFL Reachability.}

Given a weighted graph, the shortest-path CFL reachability algorithm is to find the shortest-path that conforms to the context-free grammar.
The shortest-path CFL reachability was first proposed to infer missing specifications in Android libraries~\cite{DBLP:conf/popl/BastaniAA15}.

The shortest-path CFL reachability algorithm is an extension to the basic CFL reachability algorithm.
Instead of the work list, the shortest-path CFL reachability algorithm maintains a priority queue $H$, in which an element with a higher weight is served after an element with a lower weight, and introduces a map $I$ to store the shortest path.
In the initialization stage, $H$ stores the edges with their weights in $G$ labeled with terminal symbols, and $I$ is empty.
Then, additional edges are added to $G$ according to the productions until no more edges can be added as in CFL reachability.

\subsection{Attribute Grammar}

For compiler construction, attribute grammars are used in addition to parsing to represent semantic information~\cite{DBLP:conf/waga/Knuth90}. In attribute grammars, each symbol is associated with inherited and synthesized attributes. The inherited attributes are computed in a top-down manner, while the synthesized attributes are computed in a bottom-up manner. Rules are defined to verify the compatibility of the given attributes and then compute the inherited/synthesized attributes. If the attributes are incompatible, the compiler raises an error to indicate that the input program violates semantic constraints.

\section{Approach}

\subsection{Overview}

In this subsection, we provide a brief illustration of \techname using a toy example. The details of components of \techname will be described in the next subsections.

We consider a simple programming language that contains only one assignment statement, described by the following context-free grammar:
$$\mathcal{S} \to \texttt{identifier}\ =\ \texttt{identifier}\ ;$$
The statement contains a semicolon indicating the end of the statement.

Let us consider three identifiers in this language: $x$, $y$, and $z$. Identifiers $x$ and $y$ are of the same type denoted as type $A$, while $z$ is of a different type denoted as type $B$. Supposing that only variables of the same type can be assigned to each other, $x$ and $y$ can be assigned to each other, but neither $z$ can be assigned to $x$ or $y$, nor $x$ or $y$ can be assigned to $z$.

We focus on fixing the following code snippet:
\begin{equation}
    x = z \label{equ:failprogram}
\end{equation}

Code snippet \eqref{equ:failprogram} is not compilable due to two errors. First, there is no semicolon at the end of the statement, resulting in a syntactic error that violates the context-free grammar. Second, $x$ and $z$ are of different types and cannot be assigned to each other, resulting in a semantic error that violates the typing semantic constraints.

\begin{figure}[t]
    \centering
    {
        \small
        \begin{tikzpicture}
            \node[place] (0) {$v_0$};
            \node[place,right of=0] (1) {$v_1$};
            \node[place,right of=1] (2) {$v_2$};
            \node[place,right of=2] (3) {$v_3$};
            \path[->] (0) edge [above] node {`$x$'} (1);
            \path[->] (1) edge [above] node {`$=$'} (2);
            \path[->] (2) edge [above] node {`$z$'} (3);
            \path[->] (0) edge [loop above,in=45,out=135,looseness=8,dashed] node [align=center] {`$x$', `$y$', `$z$', `$=$', `$;$'} (0);
            \path[->] (1) edge [loop above,in=45,out=135,looseness=8,dashed] node [align=center] {$\cdots$} (1);
            \path[->] (2) edge [loop above,in=45,out=135,looseness=8,dashed] node [align=center] {$\cdots$} (2);
            \path[->] (3) edge [loop above,in=45,out=135,looseness=8,dashed] node [align=center] {$\cdots$} (3);
            \path[->] (0) edge [below,in=-135,out=-45,looseness=1,dashed] node [align=center] {`$y$', `$z$', `$=$', `$;$', $\epsilon$} (1);
            \path[->] (1) edge [below,in=-135,out=-45,looseness=1,dashed] node [align=center] {$\cdots$} (2);
            \path[->] (2) edge [below,in=-135,out=-45,looseness=1,dashed] node [align=center] {$\cdots$} (3);
        \end{tikzpicture}
    }
    \caption{Modification graph for $x=z$. Multiple symbols on the one edge indicate multiple edges.}\label{fig:mod}
\end{figure}
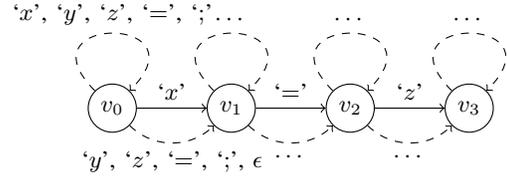

To fix compilation errors in the program, our approach, \techname, initially constructs a weighted modification graph as depicted in \figurename~\ref{fig:mod}. The graph comprises original edges and candidate modification edges. Original edges correspond to the current tokens in the code snippet and are assigned a weight of 0 (solid edges in \figurename~\ref{fig:mod}). Candidate modification edges denote possible modifications to the code snippet, including insertion (self-loops), deletion (edges from $v_i$ to $v_{i+1}$ with the empty symbol $\epsilon$), and update (edges from $v_i$ to $v_{i+1}$ with a non-empty symbol). Candidate modification edges are assigned a weight of 1 (dashed edges in \figurename~\ref{fig:mod}). After constructing the weighted modification graph, a path from the first vertex ($v_0$) to the last vertex ($v_3$) in the modification graph indicates a modified program, and the total weight of the path indicates the number of modifications required to transform the original code snippet into the modified program.

Subsequently, our approach attempts to find a compilation error fix by checking from a small number of modifications to a large number of modifications. Specifically, \techname utilizes an adapted shortest-path CFL reachability algorithm on the graph to identify a group of programs that conform to syntactic constraints with the minimum number of modifications. Thereafter, \techname checks the semantic constraints of these programs through merged attribute checking. If any of these programs satisfies semantic constraints, \techname constructs the fix and returns the result. If not, \techname employs the adapted CFL reachability algorithm again to identify another group of programs with more modifications and verifies these programs with merged attribute checking. This process continues until a fix is found.

For the CFL reachability part, we made a slight modification to the shortest-path CFL reachability algorithm by storing not only the shortest path but also other paths. This modification enables our algorithm to find fixes with more modifications. For example, in the above illustration, the CFL reachability initially finds a program with one modification, $x = z;$ (inserting a semicolon at the end of the program). In the subsequent round, CFL reachability identifies several programs with two modifications. A comprehensive description of the weighted modification graph and the CFL reachability utilized in our approach will be provided in Section~\ref{sec:syn}.

For merged attribute checking, the process of merged attribute checking involves converting semantic constraints into attributes and subsequently verifying their correctness. To apply this technique to the aforementioned code snippet, each identifier is assigned a synthesized attribute indicating its type based on the corresponding edge in the CFL reachability graph (i.e., `$x$' edges and `$y$' edges with type value $A$, and `$z$' edges with type value $B$). Then, \techname checks whether the types of the corresponding identifiers in statements match by checking their respective attributes. Further details regarding the process of merged attribute checking will be presented in Section~\ref{sec:sem}.

Attribute checking can combine branches from different locations, resulting in exponential time complexity relative to the weight of the program. To address this issue, we employ pruning and optimization techniques to scale type checking. It is important to note that since compilation error fixing is NP-hard, even with optimizations, the complete algorithm necessarily leads to exponential time complexity in the worst case, unless $\mathrm{P} = \mathrm{NP}$. Nevertheless, our experiments demonstrate that \techname can successfully identify fixes in a reasonable time. We will provide further details on the algorithm in the next subsections.

For the above example, \techname will find a fix with two modifications.
$$v_0 \xrightarrow{x} v_1 \xrightarrow{=} v_2 \xrightarrow{y} v_3 \xrightarrow{;} v_3$$

\subsection{Dealing with Syntactic Constraints}\label{sec:syn}

To find syntactically correct fixes in compilation error fixing, i.e., finding programs that conform to the context-free grammar, we define a modification graph and then reduce the problem to shortest-path CFL reachability.

For a program $p$ with $n$ tokens, we construct a weighted modification graph. The weighted modification graph is a directed graph with each edge associated with a length (weight) $l$ and a grammar symbol\footnote{A grammar symbol is a terminal symbol or a non-terminal symbol in the context-free grammar.} $t$. The weighted modification graph consists of $n+1$ vertices (i.e., $v_0, v_1, \cdots, v_n$). Each edge of the weighted modification graph is represented by $\left<v_i, v_j, t, l\right>$, where $v_i$ and $v_j$ is the original and the target vertices, $t$ is the associated symbol and $l$ is the weight of the edge.

The edges of weighted modification graph consist of four kinds of edges, one of them representing original tokens in the original program and the three others representing three kinds of modification operations. The weights of original edges are zero while the weights of other edges representing modification operations are one, so that the weight of each edge represents the number of modifications. The four kinds of edges are defined as below.
\begin{enumerate}
    \item \textit{original edge}: An original edge represents a token in the original program.
          For the $i$-th token $t_i$ in original program, we have $\left<v_{i-1}, v_i, t_i, 0\right>$ in the graph (denoted as $E_M$).
    \item \textit{insertion edge}: An insertion edge is represented by a self-loop edge labeled with the inserted symbol.
          For each vertex $v_i \in V$, edges in the form of $e_{ins}=\left<v_i,v_i,t_{ins},1\right>$ is added to $E_M$, where $t_{ins}$ is the inserted symbol before $t_i$.
    \item \textit{update edge}: An update edge is represented by an edge from vertex $v_i$ to vertex $v_{i+1}$ labeled with an update symbol.
          For each pair of vertices $v_i,v_{i+1} \in V$, edges in the form of $e_{upd} = \left<v_i,v_{i+1},t_{upd},1\right>$ is added to $E_M$, where $t_{upd}$ is a symbol used to update $t_i$ ($t_{upd} \neq t_i$).
    \item \textit{deletion edge}: A deletion edge is represented by an edge from vertex $v_i$ to vertex $v_{i+1}$ labeled $\epsilon$.
          For each pair of vertices $v_i,v_{i+1} \in V$, edge $e_{del}=\left<v_i,v_{i+1},\epsilon,1\right>$ is added to $E_M$.
\end{enumerate}

According to the above construction, given a path from $v_0$ to $v_n$ in $G$ (denoted as $P$), if the weight of the path is $k$, $P$ corresponds to $k$ modifications to $p$.

By using shortest-path CFL reachability on the modification graph, we can generate CFL reachability graphs that contain the syntactic structures of both the original and modified programs. On the CFL reachability graph, an edge in the form of $e_s=\left<v_0, v_n, \mathcal S, k\right>$, where $v_0$ is the first vertex, $v_n$ is the last vertex and $\mathcal S$ is the start symbol, represents one or many syntactically correct programs with $k$ modifications. We can construct syntactically correct programs from each $e_s$ by recursively expanding productions of non-terminals. So, we refer to these $e_s$ edges as root edges in the following discussion.

Using the shortest-path CFL reachability algorithm, we can enumerate root edges from shorter to longer and find syntactic fixes in order of increasing length. In the next subsection, we will further apply syntactic constraints to these fixes.

\subsection{Dealing with Semantic Constraints}\label{sec:sem}

After obtaining root edges, in this subsection, we further utilize attribute checking on the reachability graph to identify semantic error fixes. This is achieved by encoding semantic constraints into attributes on the symbols, allowing for a more precise and effective error correction process.

\subsubsection{Encoding Semantic Constraints into Attributes}

We encode semantic constraints into attributes, which will be checked in the reachability graph to find shortest-path fixes.
While global identifier tables are typically used during parsing for a single program, when searching for semantically correct programs on a CFL reachability graph, multiple potential fixes can exist, and the global identifier tables for different reachability paths may vary.
Additionally, for the attribute checking algorithm to work effectively in merging and pruning, the attributes must be immutable, hashable, and comparable.
To overcome these challenges, we develop a new method for encoding semantic constraints into attributes without relying on global identifier tables.

\paragraph{Declaration-Use Constraints}
Declaration-use constraints are a fundamental aspect of programming languages, where the compiler detects an error when an undeclared identifier is used.
In our approach, we encode declaration-use constraints in the reachability graph by introducing an inherited attribute to store declaration information, along with identifier selectors to determine whether an identifier can be used. Specifically, for each grammar symbol (i.e., a terminal or a non-terminal in the context-free grammar), we introduce an inherited attribute $idtab$ to store the identifiers defined before this symbol. For an identifier terminal $x$, we introduce another inherited attribute $sel$ to contain possible selections for the identifier. Both $idtab$ and $sel$ utilize an immutable map data structure to store the mapping from identifier names to their types and an immutable list data structure to store possible identifier names, respectively.

\paragraph{Expression Type Constraints}
In most programming languages, expressions are associated with types, and operations between expressions must satisfy specific type constraints. These are known as expression type constraints in programming languages. To handle these constraints, we use a synthesized attribute $type$ to store the type of the expression. Computation of a new synthesized attribute checks whether the operation between sub-expressions is valid, and if so returns the type of the result. For example, for sum of two primaries $p = x + y$, the computation checks whether the type of $x$ and $y$ are compatible in an addition operation and returns the type of $a+b$ if $a$ and $b$ are compatible. The computation of this attribute varies depending on the programming language, so we use specific data structures for different languages. For function invocations, we introduce another inherited attribute to store an immutable list of types of arguments, which will help check the expression types of parameters.

\paragraph{Other Constraints}
In certain programming languages, there may be additional types of constraints beyond declaration-use and expression type constraints. For example, in C, the keywords ``break'' and ``continue'' can only be used within loop or switch statements, which can be checked by introducing a new inherited attribute to store information about the enclosing loop or switch statement. Similar to the constraints mentioned earlier, these other constraints can also be encoded into attributes similarly in compiler construction.

\subsubsection{Merged Attribute Checking on Reachability Graph}

Above, we have encoded semantic constraints into attributes. In this subsection, we further describe how to check these attributes on the CFL reachability graph.

Before checking attributes, we should first transform the calculation of attributes into a normalized grammar form, because, in CFL reachability, the context-free grammar used are in a normalized form, where the right-hand side of each production rule contains at most two symbols.
This transformation is easy to implement, by expanding the synthesized attributes into a list.
After normalization, the synthesized attributes of each symbol are converted into an array of synthesized attributes in new rules and then computed according to the original grammars. If inherited attributes exist in the new rules, they are copied and computed based on the original inherited attributes.

Next, we check computation of attributes for a normalized grammar in the CFL reachability graph.
To explain how \techname checks attributes on the reachability graph, we first describe the process of computing and checking attributes in recursive descent compilers. Consider a production rule $A \to BC$. During parsing of attribute grammars, the inherited attribute of symbol $X$ is denoted by $I_X$, and the synthesized attribute of symbol $X$ is denoted by $S_X$. The attributes of $A, B$, and $C$ are computed in the following sequence:
\begin{algorithmic}[1]
    \STATE Compute $I_B$ from $I_A$
    \STATE Recursively descent to $B$ and compute $S_B$ from $I_B$
    \STATE Compute $I_C$ from $I_A$ and $S_B$
    \STATE Recursively descent to $C$ and compute $S_C$ from $I_C$
    \STATE Compute $S_A$ from $I_A$, $S_B$, and $S_C$
\end{algorithmic}

To check inherited and synthesized attributes on the reachability graph, a similar recursive descent approach can be used as in compilers. However, there is a key difference: in a compiler, there is only one parse tree and each symbol has only one value for its synthesized attribute, while in the CFL reachability of a weighted modification graph, there can be multiple parse trees, and a non-terminal symbol on an edge may represent multiple paths of reduction. Therefore, it is necessary to maintain a set of inherited and synthesized attribute values for an edge on the reachability graph, and for each path of reduction, \techname computes attributes and inserts them into the appropriate sets. Specifically, for the production rule $A\to BC$ in the previous example, \techname uses a three-level loop as shown in \figurename~\ref{fig:threeloop}.
This three-level loop iterates over all possible reductions of $A \to BC$, all possible synthesized attribute values of $B$, and all possible synthesized attribute values of $C$, and computes the synthesized attribute value of $A$ for each possible reduction.

\begin{figure}[t]
    \small
    \centering
    \begin{framedsized}[.45\textwidth]
        \begin{algorithmic}[1]
            \FOR{each reduction of $A \to BC$}
            \STATE Compute $I_B$ from $I_A$
            \STATE Recursively descent to $B$ and compute the set of $S_B$ from $I_B$
            \FOR{each element of $S_B$}
            \STATE Compute $I_C$ from $I_A$ and $S_B$
            \STATE Recursively descent to $C$ and compute the set of $S_C$ from $I_C$
            \FOR{each element of $S_C$}
            \STATE Compute $S_C$ from $I_A$, $S_B$, and $S_C$
            \STATE Add $S_A$ into the synthesized attribute set.
            \ENDFOR
            \ENDFOR
            \ENDFOR
        \end{algorithmic}
    \end{framedsized}
    \caption{Three-level loop for attribute checking.}
    \label{fig:threeloop}
\end{figure}

The algorithm described above has the potential to produce an exponential number of attribute values when combining attribute values on different paths. To mitigate this issue, we have implemented several optimizations in our approach focusing on proper memoization, merging, and early pruning during the checking process. These optimizations make our algorithm applicable for real-world applications. We will further discuss these optimizations in the next subsection.


\subsection{Algorithm for Fixing Compilation Errors}

In \figurename~\ref{alg:compiling}, we present the algorithm for \techname\footnote{Due to spatial constraints, we have included the algorithms for \texttt{CFLReachability} and \texttt{CheckAttr} in the appendices.
}. The algorithm begins by constructing a weighted modification graph. Next, it uses the \texttt{CFLReachability} procedure to find the shortest root edge and then checks the semantic correctness using the \texttt{CheckAttr} procedure. If the program is semantically correct, it is constructed from the graph and returned. If the program is not semantically correct, the algorithm uses the \texttt{CFLReachability} procedure again to find the next longer root edge and repeats the process of checking the root edge using \texttt{CheckAttr}. This process is repeated until a syntactically and semantically correct program is found.

\begin{figure}[t]
    \centering
    \small
    \begin{framedsized}[.45\textwidth]
        \begin{algorithmic}[1]
            \REQUIRE $p$, a program to be fixed
            \STATE $G$ = ConstructModificationGraph($p$)
            \WHILE{$e_s$ = CFLReachability($G$) not empty}
            \IF{CheckAttr($e_s$, $I_s$) is not empty}
            \RETURN ConstructResult($e_s$, $I_s$)
            \ENDIF
            \ENDWHILE
        \end{algorithmic}
    \end{framedsized}
    \caption{Algorithm for compilation error fixing.}\label{alg:compiling}
\end{figure}

To construct the CFL reachability graph in \techname, we use an adapted shortest-path CFL reachability algorithm that collects all possible paths in the weighted modification graph. Unlike the original algorithm that only stores one shortest edge, our approach stores all CFL reachability edges since different edges may lead to different attributes during subsequent attribute checking. Similar to the original CFL reachability algorithm, we maintain a priority queue as a working queue to save all edges that need further processing. In the queue, edges with higher weight are processed after edges with lower weight, which is the same as the original shortest-path CFL reachability. However, our algorithm differs from the original by prioritizing edges with the start symbol later than other edges when they have the same weight. This ensures that all children of an edge labeled with the start symbol are obtained before attribute checking, thereby preserving the invariance of sub-problems in the dynamic programming of attribute checking, as discussed in the following paragraphs.

Next, we discuss the design of \texttt{CheckAttr}. In the following discussion, we assume that there is only one inherited attribute and only one synthesized attribute with each symbol, which can be generalized to situations where multiple inherited or synthesized attributes exist. Inspired by the top-down approach of dynamic programming with memoization used in recursive descent parsers in compiler construction~\cite{recursive-descent-parsing}, we design the \texttt{CheckAttr} algorithm as follows. Given an edge in the CFL reachability graph and an inherited attribute value, \texttt{CheckAttr} returns a set of all possible synthesized attribute values. In contrast to recursive descent parsers that construct one parse tree during parsing, \techname computes attribute values for one or more parse trees that have already been constructed during CFL reachability and returns a set of possible values for each synthesized attribute.

The following optimizations have been implemented in \texttt{CheckAttr}.

\textbf{Pruning.} To improve efficiency, we prune edges as early as possible during attribute checking when any attribute constraint is not satisfied.

\textbf{Merging.} We have observed that different program fragments can have the same semantic meaning. For instance, two different expressions can be of the same type, indicating that programs with either expression will either both compile or both not compile. Therefore, after checking two expressions, we merge them before checking the remaining program. To accomplish this, we merge attributes with identical values during the three-level loop of attribute computation.

\textbf{Memoization.} Since only a few modifications are made in the program, most remaining program fragments are the same. As a result, a reachable edge in the graph will be checked multiple times, and during different rounds of attribute checking, most remaining program fragments remain the same. To prevent duplicate computation, we use memoization. We maintain a global map to store the function's output, which ensures that the same input always returns the same output. This eliminates redundant computations, thereby enhancing the algorithm's efficiency. To maintain the invariance of the synthesized attribute value set, a priority queue is used in \texttt{CFLReachability}. The queue serves root edges $e_s$ with weight $p$ later than all other edges shorter than or equal to $p$ weight. As a result, when \texttt{CheckAttr} is invoked, all descendants of $e_s$ are already constructed, and no new children edges of $e_A$ are added after \texttt{CheckAttr}$(e_A, I_A)$ is executed.


%% file: chaps/06-experiments.tex
\section{Experiments}

We conducted an experimental study to evaluate the effectiveness of \techname in fixing compilation errors. Our implementation adopts a two-layer architecture: the lower layer includes an algorithm for CFL reachability with attribute checking, while the upper layer contains a language-specific fixer. By mapping the context-free grammar and attribute checking rules to the underlying algorithm, we can easily build a bug fixer for a new language.

To evaluate the effectiveness of \techname, we conducted experiments on programming languages with static typing. We implemented error fixers for two languages: Middleweight Java, an object-oriented language, and C, a procedural language. Type checking in Middleweight Java and C focuses on different relationships: Middleweight Java type checking examines function call relationships between user-defined classes, while C type checking concentrates on arithmetic and conversion relationships between basic data types. We used Middleweight Java to demonstrate the effectiveness of \techname in capturing external class definition relations and used C to demonstrate the effectiveness of \techname in capturing internal correctness of function bodies.

Since Middleweight Java is not widely used in practice, we generated a set of random compilable programs in this language and introduced compilation errors using a series of mutation operators. For C, we used the popular DeepFix dataset~\cite{DBLP:conf/aaai/GuptaPKS17}, which contains real-world student C programs with compilation errors, for evaluation. Our implementation focuses on fixing errors in function bodies for a given language. Therefore, our experiments focus on evaluating the effectiveness of \techname for fixing function body errors.

\subsection{Middleweight Java}

Middleweight Java is a lightweight subset of the Java programming language~\cite{bierman2003mj} that includes important features such as objects, inheritance, and polymorphism. In this study, we implemented and evaluated \techname to fix syntactic and/or semantic compilation errors in method bodies written in Middleweight Java.

The original Middleweight Java grammar~\cite{bierman2003mj} is designed for abstract syntax, and thus, it does not consider operator precedence or parentheses. To parse a concrete program using this grammar, we adapted it to include bracket expressions and a new symbol, primary $p$, which handles operator precedence.

Since Middleweight Java is not widely used in practice, to evaluate \techname, we generated a set of random compilable Middleweight Java programs and introduced compilation errors through a series of mutation operators.

\subsubsection{Compilable Middleweight Java Program Generation}

To ensure that a generated Middleweight Java program is compilable, we generate both the method body and all the class definitions. The following steps outline our detailed generation process: (1) We generate a random number of classes with unique names. (2) We randomly assign inheritance relationships among these classes and ensure that there is no cycle in the inheritance graph and all classes are inherited from the \texttt{Object} class. (3) We generate a constructor, a random number of methods with random parameters, and a random number of fields for each class. (4) We generate a compilable method body using top-down derivation according to the context-free grammar, and ensure that semantic constraints are satisfied during the derivation.

In the last step, we use a recursive descent generation process. When performing generation for a symbol, we randomly select a production rule associated with the symbol and then call subroutines for the symbols on the right-hand side of the production rule. We also pass down the variable table and expression types to check whether the generation is feasible. If a call to a generation subroutine fails, such as when attempting to generate a local variable with an empty variable table, we trigger a backtracking operation to select a new random production rule in some outer subroutines.

\begin{figure}[t]
    \centering
    \subfloat[Number of tokens]{\includegraphics[width=.24\textwidth]{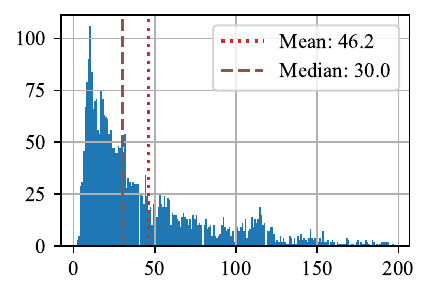}}
    \hfil
    \subfloat[Number of lines]{\includegraphics[width=.24\textwidth]{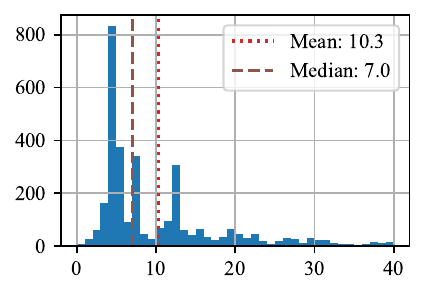}}
    \caption{The length distribution of Middleweight Java method bodies.}\label{fig:length}
\end{figure}

The generation of 1,000 Middleweight Java programs is completed within a few seconds. In the following, we present statistical information about the generated programs. \figurename~\ref{fig:length} shows the length distribution of method bodies in terms of both the number of tokens and the number of lines, represented as a histogram. The average number of tokens is 45.2, and the average number of lines is 10.3. These statistics indicate that the generated programs are of reasonable size for method bodies.

\subsubsection{Middleweight Java Program Mutation}

To evaluate the effectiveness of our approach to compilation error fixing, we need a set of programs with compilation errors to use as test cases. To create such programs, we applied mutations to generate programs with syntactic or semantic errors.

In Middleweight Java, identifiers and other tokens behave differently. For example, replacing punctuation marks or keywords is more likely to lead to syntactic errors, while replacing identifiers is more likely to cause semantic errors.

We designed eight mutation operators (see Table~\ref{tab:mutationops}) that include three basic operations (insertion, deletion, and replacement) and a special case of insertion, i.e., duplication. The mutation operators were applied to our generated compilable programs to produce new programs with compilation errors. We used a combination of syntactic and semantic error mutations to create a diverse set of test cases for our experiments.

\begin{table}[t]
    \centering
    \caption{Mutation Operators}
    \label{tab:mutationops}
    \begin{tabular}{cl}
        \toprule
        ID  & Operator description                         \\
        \midrule
        M.1 & Inserting a punctuation mark or a keyword.   \\
        M.2 & Deleting a punctuation mark or a keyword.    \\
        M.3 & Duplicating a punctuation mark or a keyword. \\
        M.4 & Replacing a punctuation mark or a keyword.   \\
        M.5 & Inserting an identifier.                     \\
        M.6 & Deleting an identifier.                      \\
        M.7 & Duplicating an identifier.                   \\
        M.8 & Replacing an identifier.                     \\
        \bottomrule
    \end{tabular}
\end{table}

For each generated program, we produced three groups of mutants by applying the above mutation operators, i.e., a group of mutated programs with syntactic errors (denoted as \textit{Group$_{syn}$}), a group of mutated programs with semantic errors (denoted as \textit{Group$_{sem}$}), and a group of mutated programs mixing both syntactic and semantic errors (denoted as \textit{Group$_{mix}$}).
Specifically, we applied M.1$\sim$4 to produce \textit{Group$_{syn}$} since the four mutation operators are much more likely to produce syntactic errors.
We applied M.8 to produce \textit{Group$_{sem}$} since identifier replacement tends to cause semantic errors.
We applied all the eight mutation operators to produce \textit{Group$_{mix}$}.
During the process of generating a mutated program, we first determined the mutation order as 1$\sim$8.
Then, for each mutation order $i$ (1 $\leq$ i $\leq$ 8) we randomly selected a mutation operator from the corresponding set of mutation operators (such as M.1$\sim$4 for \textit{Group$_{syn}$}) to mutate the program (i.e., the compilable method body), and repeated this step $i$ times\footnote{We introduced $i$ errors to the method body by mutating the program $i$ times, which means that the errors can be fixed (i.e., find a compilable method body) with fewer than or equal to $i$ modifications.}.
In this way, we obtained 8 mutants for each group with regard to a program.
In total, we obtained 2400 mutated programs, including 800 mutants in \textit{Group$_{syn}$}, and 800 mutants in \textit{Group$_{sem}$}, 800 mutants in \textit{Group$_{mix}$}.

\subsection{C Programs}\label{sec:cprocessor}

In our evaluation on C programs, we used a dataset of 6,978 student C programs with compilation errors that was previously used to evaluate DeepFix~\cite{DBLP:conf/aaai/GuptaPKS17}.
To enable type checking, we implemented a subset of C that covers all features used by students.
The context-free grammar of this simplified C contains 29 non-terminal symbols and 132 production rules, which is more complex than the grammar of Middleweight Java.
Additionally, we designed type checking rules based on the C programming language to support the determination of left values and right values, pointers, expressions, and function invocations.

\begin{figure}[t]
    \centering
    \subfloat[Number of tokens]{\includegraphics[width=.24\textwidth]{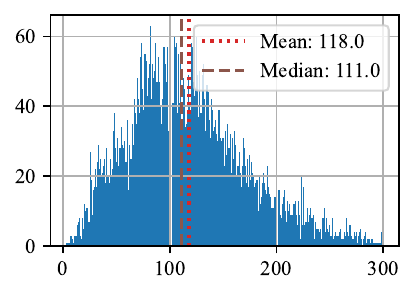}}
    \hfil
    \subfloat[Number of lines]{\includegraphics[width=.24\textwidth]{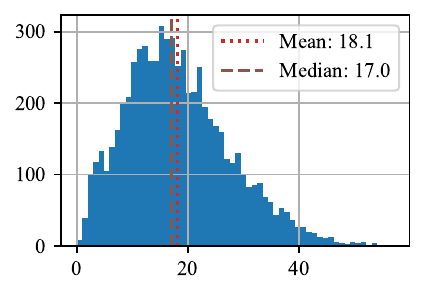}}
    \caption{The length distribution of C method bodies.}\label{fig:clength}
\end{figure}

Since our C fixer does not recognize macros, we pre-processed the given C programs using a C pre-processor. To identify the function bodies, we used a processor that pairs the opening brackets (i.e., ``\{'') and the closing brackets (i.e., ``\}''). If we encountered more opening brackets than closing brackets, we attempted to add a closing bracket at the end of the program. If a program failed the pre-processing step or had compilation errors outside the function bodies, we skipped it. We were left with 6,536 function bodies to fix. The length distribution of the function bodies is presented in \figurename~\ref{fig:clength}.

\subsection{Results}

Based on these mutated Middleweight Java programs and the student C programs with compilation errors, we applied \techname to fix them in order to evaluate the performance of \techname.
Our study was conducted on a workstation with Intel(R) Xeon(R) E5-2680 v4 CPUs. We set the time limit to 600 seconds and the memory limit to 15 GB for fixing a mutated program. It is important to note that we only ran one thread, even though our platform has multicore CPUs, so our evaluation results show single CPU times.

\subsubsection{Middleweight Java}

\begin{figure}[t]
    \centering
    \includegraphics[width=.4\textwidth]{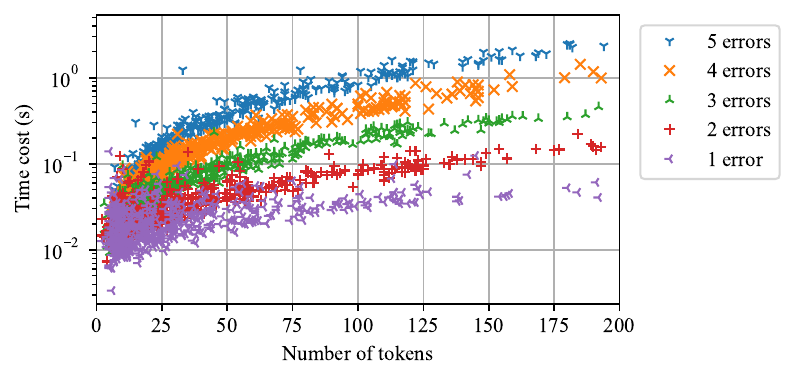}
    \caption{Middleweight Java fixing time scatter diagram.}\label{fig:time-to-totkens}
\end{figure}

\begin{figure}[t]
    \centering
    \subfloat[1 error]{\includegraphics[width=.24\textwidth]{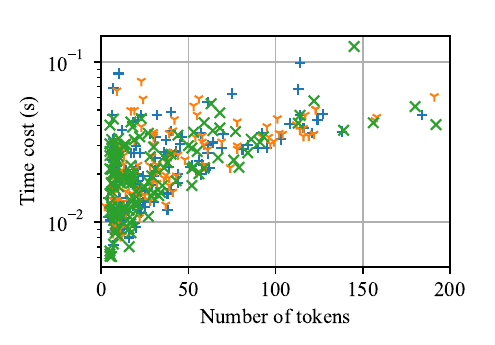}}
    \hfil
    \subfloat[3 errors]{\includegraphics[width=.24\textwidth]{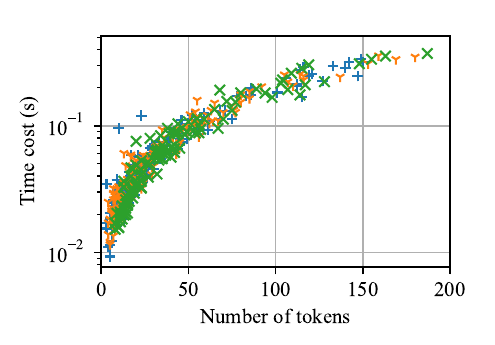}}\\
    \subfloat[5 errors]{\includegraphics[width=.24\textwidth]{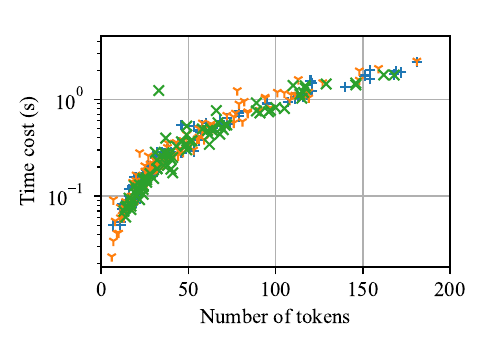}}
    \hfil
    \subfloat[Labels]{\includegraphics[width=.24\textwidth]{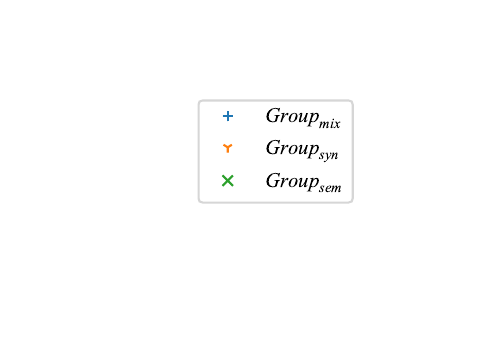}}
    \caption{Middleweight Java fixing time scatter diagram by type.}\label{fig:time-to-kind}
\end{figure}

We observe that \techname is able to successfully fix all the compilation errors (including both syntactic and semantic errors).
The results demonstrate the effectiveness of \techname.

We further analyze the factors affecting the efficiency of \techname, whose results are shown in \figurename~\ref{fig:time-to-totkens}.
In this figure, the x-axis represents the length of a program in terms of the number of tokens, the y-axis represents the time cost spent on fixing a program in the logarithmic scale, and different marks represent different numbers of compilation errors in a program.
From this figure, regardless of the number of compilation errors in a program, the growth of the time costs becomes slower in the logarithmic scale with the number of tokens increasing, indicating that the growth of the time costs with the program length is less than exponential growth.
Regarding the same length of programs, the growth of the time costs is stable in the logarithmic scale with the number of errors increasing, indicating that the growth of the time costs with the number of errors is almost exponential.
That is, the time cost spent on fixing a program is more correlated with the number of errors in the program than the program length in terms of the number of tokens.

\figurename~\ref{fig:time-to-kind} further analyzes the relationship between the time costs spent on fixing a program and different mutant groups, where different marks represent different groups of mutated programs.
We find that there are very similar trends for the three groups regardless of the number of errors in a program, indicating that the efficiency of \techname is not affected by different types of compilation errors.
Also, we find that \techname spends less than 10 seconds on fixing a program around 200 tokens with 5 errors. The results demonstrate the efficiency of \techname.

\subsubsection{C}

\begin{table}[t]
    \caption{C function body fixing results}
    \label{tab:cresult}
    \begin{center}
        \begin{tabular}{lcc}
            \toprule
            \textbf{Outcome}      & \textbf{Count} & \textbf{Percentage} \\
            \midrule
            Success               & 5,888          & 90.0\%              \\
            Memory Limit Exceeded & 627            & 9.6\%               \\
            Time Limit Exceeded   & 10             & 0.2\%               \\
            Other                 & 11             & 0.2\%               \\
            Total                 & 6,536          & 100.0\%             \\
            \bottomrule
        \end{tabular}
    \end{center}
\end{table}

Table~\ref{tab:cresult} presents the fixing outcomes of \techname on the C function bodies. In addressing compilation errors, \techname achieves a commendable success rate of 90.0\% (5,888 out of 6,536) within time and memory limit.
The majority of failures (627 instances) are attributed to memory limit exceedance, which is mainly from the memorization mechanism utilized in our algorithm.
A subset results from time limit exceedance (10 instances), attributable to the time complexity of the compilation error resolution process.
A few other failures (11 instances) arise from internal macros employed by compilers, which our implementation does not support. For instance, in GCC, the macro ``\texttt{isprint}'' is defined in a complex manner, leading to inconsistency between the attribute grammar and the compiler's behavior.

\begin{figure}[t]
    \centering
    \includegraphics[width=.4\textwidth]{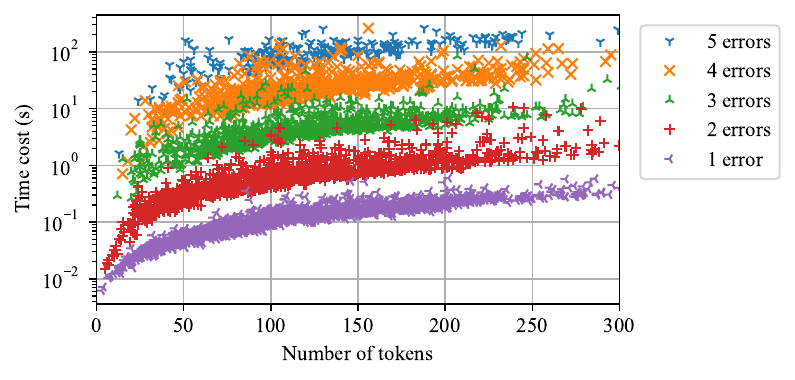}
    \caption{C fixing time scatter diagram.}\label{fig:time-to-totkens-c}
\end{figure}

\figurename~\ref{fig:time-to-totkens-c} demonstrates the time consumption of token numbers for varying numbers of modifications. \techname can fix a function body within 10 seconds with up to 3 modifications, which is acceptable in application. Similar to Middleweight Java, the increase in time costs remains steady on a logarithmic scale as the number of errors grows, suggesting that the time costs rise almost exponentially with the number of errors.

\begin{figure}[t]
    \centering
    \includegraphics[width=.2\textwidth]{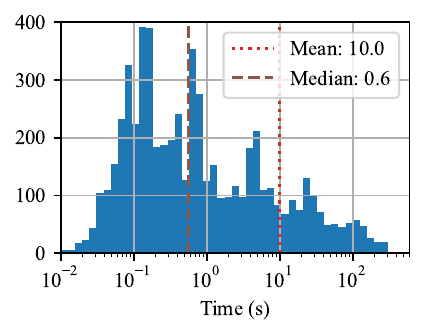}
    \caption{C fixing time distribution.}\label{fig:time-c}
\end{figure}

The distribution of fixing times for all fixed function bodies is depicted in \figurename~\ref{fig:time-c}. Our analysis reveals that the median time for fixing a function body is only 0.6 seconds, indicating that the majority of fixes are completed quickly. Specifically, 62.4\% of programs can be fixed within 1 second, and 84.2\% of programs can be fixed within 10 seconds. These results demonstrate that the time cost associated with fixing compilation errors is acceptable for real-world applications.

To compare \techname with existing methods, we also computed the fixing rate on the original dataset. \techname considers a program to be fixed only if three conditions are met: 1) our processor\footnote{See Section~\ref{sec:cprocessor}} successfully detects function bodies in the program, 2) the program compiles after the function bodies are removed, and 3) \techname successfully resolves all errors in each function body. We note that if our processor fails or the program is incorrect outside the function bodies, we consider that \techname fails to fix the program. Table~\ref{tab:baseline} displays the results of \techname and several other neural network-based approaches.
The fixing rates of the existing approaches are reported as per their respective papers, and we did not impose any additional time or memory constraints on them.
\techname achieves a success rate of 83.5\% (5,757 out of 6,978)\footnote{A C program contains one or more function bodies, so the number of fixed programs is less than the number of fixed function bodies.} for code with compilation errors, surpassing all existing approaches. In the previous paragraph, we examined the running time of \techname. Based on the distribution of our results, it is evident that, for most instances, \techname is on par with the running time of the neural network approaches, which typically takes only a few fractions of a second or a few seconds to complete. This indicates that \techname is an efficient and effective approach for fixing compilation errors.

We conduct a more detailed analysis of the instances in which \techname encounters failures in DeepFix dataset.
Our examination reveals that 507 (7.3\%) instances are attributed to processor failures, while 64 (0.1\%) instances result from programs that do not compile even after the removal of function bodies. These failures stem from erroneous structures outside function bodies, including improperly formed function declarations, malformed global variable declarations, and incorrect usage of brackets, among other issues. The remaining 650 (9.3\%) instances are associated with situations where \techname is unable to fix the function bodies, which has been analyzed in Table~\ref{tab:cresult} and preceding paragraphs.

\begin{table}[t]
    \caption{Comparison of Fixing Rate}
    \label{tab:baseline}
    \begin{center}
        \begin{tabular}{lc}
            \toprule
            \textbf{Approach}                                    & \textbf{Fixing Rate} \\
            \midrule
            DeepFix \cite{DBLP:conf/aaai/GuptaPKS17}             & 33.4\%               \\
            DrRepair \cite{DBLP:conf/icml/YasunagaL20}           & 66.1\%               \\
            BIFI \cite{DBLP:conf/icml/YasunagaL21}               & 71.7\%               \\
            TransRepair \cite{DBLP:journals/corr/abs-2210-03986} & 68.5\%               \\
            \textbf{\techname}                                   & \textbf{82.5\%}      \\
            \bottomrule
        \end{tabular}
    \end{center}
\end{table}

\section{Discussion}

The inherent NP-hardness of the compile error fixing problem highlights a significant challenge in the scalability of \techname.
In response, our algorithm has been crafted to mitigate time-related concerns.
As discussed further in the preceding sections and supported by the experimental findings, the time complexity is predominantly exponential in relation to the count of modifications required to rectify a given program.
Real-world scenarios reveal that humans typically spend several seconds to minutes addressing compilation errors~\cite{DBLP:conf/icse/SeoSEAB14}.
Comparatively, the time invested by \techname in error resolution for most function bodies in languages such as Java or student-written C programs is strikingly similar to the time expended by human counterparts.

For more complex programming languages or more complex programs, we need to design more efficient algorithms to reduce the time consumption of \techname.
Additionally, \techname finds a compilation error fix with minimum number of modifications. In some cases, there are several possible fixes with the same number of modifications. It is interesting to further use some heuristics or statistical models to guide the searching process, improve the efficiency of the algorithm and find the fix that is the most similar to the human fix. This is our future work.

%% file: chaps/07-related-work.tex
\section{Related Work}

\textit{Automatic Program Repair.} Automated program repair (APR) is a task to fix unexpected behavior in programs~\cite{DBLP:conf/icse/GazzolaMM18,DBLP:journals/cacm/GouesPR19,DBLP:journals/chinaf/CuiFCCZLL22}, i.e., the programs are compilable, but some functionality is wrong. Recently, many search-based~\cite{DBLP:conf/icse/QiMLDW14,DBLP:conf/issta/MartinezM16}, semantic-based~\cite{DBLP:conf/wcre/LiuK0B19,DBLP:journals/tosem/GaoWDJXR21}, pattern-based~\cite{DBLP:conf/kbse/GhanbariZ19,DBLP:journals/pacmpl/BaderSP019}, and deep-learning-based~\cite{DBLP:conf/kbse/SahaLYP17,DBLP:conf/issta/LutellierPPLW020,DBLP:conf/sigsoft/ZhuSXZY0Z21} APR techniques have been proposed. These methods are developed to fix functionality defects in software and usually require compilable programs.\nocite{actionablerepairxiong}

\textit{Syntactic Error Fixing and Error Recovery.} In 1972, Aho et al.~\cite{DBLP:journals/siamcomp/AhoP72} proposed an error-correction parser to find a fix for context-free grammars with minimum modifications. Later, more work focused on syntactic error recovery. We divide them into two groups: those for LL parsing~\cite{%
DBLP:journals/acta/FischerMQ80,%
DBLP:journals/cl/HammondR84,%
DBLP:journals/acta/FischerM92,%
DBLP:conf/pldi/ParrF11%
} and those for LR parsing~\cite{%
DBLP:journals/ijpp/BarnardH82,%
DBLP:journals/cl/HammondR84,%
DBLP:conf/pldi/Cormack89,%
DBLP:journals/toplas/McKenzieYV95,%
DBLP:journals/toplas/KimC01,%
DBLP:journals/toplas/CorchueloPRT02%
}. LL(*)~\cite{DBLP:conf/pldi/ParrF11} is the parsing strategy used in Antlr, which is a powerful parser generator. To produce more useful error messages, LL(*) takes into account more of the surrounding code as the context by using an increasing look-ahead scope of instead of a fixed look-ahead scope. Merr~\cite{DBLP:journals/toplas/Jeffery03} enables useful syntactic error messages in LR-based compilers by mapping parse states and input tokens to error messages. Recently, Seq2Parse~\cite{seq2parse} combines symbolic error correcting parsers and neural networks to repair parse errors effectively. These methods focus on only syntactic errors for context-free grammars.

\textit{Compilation Error Fixing.} Recently, some datasets in Java and C for compilation error fixing were released~\cite{DBLP:conf/sigcse/BrownKMU14,DBLP:journals/corr/BhatiaS16}. Several techniques~\cite{%
DBLP:conf/aaai/GuptaPKS17,%
DBLP:conf/icse/AhmedKKKG18,%
DBLP:conf/wcre/SantosCPHA18,%
DBLP:conf/aaai/GuptaKS19,%
DBLP:conf/icml/YasunagaL20,%
DBLP:conf/icml/YasunagaL21,%
DBLP:journals/corr/abs-2210-03986%
} addressed the problem of compilation error fixing targeting student programs. The state-of-the-art approach Break-It-Fix-It~\cite{DBLP:conf/icml/YasunagaL21} designed a graph neural network model to process the source code and diagnostic feedback information. These techniques use statistical models to predict the repair results. Different from these approaches, we model the problem of compilation error fixing as the problem of finding compilable programs with minimum modifications and propose a complete algorithm to solve the problem.

%% file: chaps/08-conclusion.tex
\section{Conclusion}

In this paper, we consider the problem of compilation error fixing with minimum modifications. We show the NP-hardness of this problem and present a complete algorithm for this problem via shortest-path CFL reachability with attribute checking. We also conducted an experimental study to evaluate \techname on a number of generated Middleweight Java programs and real-world student erroneous C programs, demonstrating its effectiveness.

%% file: chaps/09-algorithm.tex
\appendix[Algorithms for \texttt{CFLReachability} and \texttt{CheckAttr}]

We briefly describe \texttt{CFLReachability} and \texttt{CheckAttr} in \figurename~\ref{fig:alg-cflreachability} and \figurename~\ref{fig:alg-checkattr}.

\begin{figure*}[t]
    \centering
    \small
    \begin{framedsized}[.9\textwidth]
        \begin{algorithmic}[1]
            \PROCEDURE{CFLReachability}{$G$, $H$}
            \STATE\COMMENT{$G$ is the weighted modification graph}
            \STATE\COMMENT{$H$ is the working queue with priority, which stores edges to be processed}
            \IF{first time to call \texttt{CFLReachability}}
            \FOR{each rule $A \to \epsilon$, each vertex $v_i$}
            \STATE add $\left<v_i, v_i, A, 0\right>$ to $G$
            \ENDFOR
            \FOR{each edge $e$ $\in$ $G$}
            \STATE $H$.push($e$)
            \ENDFOR
            \ENDIF
            \WHILE{$H$ not empty}
            \STATE $e$ $=$ $H$.pop()
            \STATE $A$ $=$ $e$.label
            \IF{$A$ is the start symbol and $e$ begins at $v_0$ and ends at $v_n$}
            \RETURN $e$
            \ENDIF
            \STATE \texttt{try\_production\_1}($e$, G, $H$)
            \STATE \texttt{try\_production\_2\_left}($e$, G, $H$)
            \STATE \texttt{try\_production\_2\_right}($e$, G, $H$)
            \ENDWHILE
            \ENDPROCEDURE
        \end{algorithmic}
    \end{framedsized}
    \caption{The \texttt{CFLReachability} algorithm.}
    \label{fig:alg-cflreachability}
\end{figure*}

\figurename~\ref{fig:alg-cflreachability} shows the \texttt{CFLReachability} algorithm.
The procedure \texttt{CFLReachability} operates similar to the shortest-path CFL reachability algorithm~\cite{DBLP:conf/popl/BastaniAA15}.
The procedure \texttt{CFLReachability} operates on the weighted modification graph, referred to as $G$.
Additionally, \texttt{CFLReachability} utilizes a priority-based working queue, denoted as $H$, to store edges to be processed.
The priority queue $H$ is ordered by the weight of edges, where edges with lower weights are processed first.
For edges with the same weight, root edges are processed last in the queue $H$ to ensure the invariance of sub-problems in the dynamic programming.
The queue $H$ is implemented as a heap with a customized comparator.

When initially invoked, \texttt{CFLReachability} carries out a preparatory step, aiming to handle $A \to \epsilon$ production rules, as shown in \figurename~\ref{fig:gen-0}.
Subsequently, the edges within $G$ are enqueued into $H$ to initialize the processing sequence.

The core of the algorithm is a looping construct that continues until the working queue $H$ becomes empty.
Within each iteration, the algorithm dequeues an edge $e$ from $H$, capturing its associated label $A$.
If $A$ corresponds to the start symbol and the edge begins at the first vertex and ends at the last vertex, the algorithm promptly terminates, signifying the attainment of a valid path.
Otherwise, \texttt{CFLReachability} systematically explores potential productions to reduce the current path.
It does so by invoking three distinct subroutines: \texttt{try\_production\_1}, \texttt{try\_production\_2\_left}, and \texttt{try\_production\_2\_right}.
These subroutines extend the exploration process, utilizing the edge's information and the graph $G$ to attempt various production possibilities, as shown in \figurename~\ref{fig:gen-1} and \figurename~\ref{fig:gen-2}.

Different from original shortest-path CFL reachability algorithm, when inserting an edge with a higher weight into the graph, \texttt{CFLReachability} does not remove the edge with the same label and the same end vertex from the graph to avoid the loss of potential solutions. 

Through this algorithm, \texttt{CFLReachability} endeavors to uncover viable paths that contain syntactically correct programs.

\begin{figure*}[p]
    \centering
    \small
    \begin{framedsized}[.9\textwidth]
        \begin{algorithmic}[1]
            \PROCEDURE{CheckAttr}{$e_A$, $I_A$, $\mathbb M$}
            \STATE\COMMENT{$e_A$ is the edge to be checked}
            \STATE\COMMENT{$I_A$ is the inherited attribute value}
            \STATE\COMMENT{$\mathbb M$ is the memoization map}
            \STATE\COMMENT{denote the symbol of $e_A$ as $A$}
            \IF{$\left(e_A, I_A\right)$ is in $\mathbb M$}
            \RETURN $\mathbb M\left[\left(e_A, I_A\right)\right]$
            \ENDIF
            \STATE construct empty list $\mathbb S_A$ to store results
            \STATE $\mathbb M\left[\left(e, I\right)\right] = \mathbb S_A$
            \IF{$A$ is an edge with terminal symbol}
            \STATE $S_A$ $=$ \texttt{process\_terminal}($A$, $I_A$)
            \IF{$S_A$ is not \textbf{null}}
            \STATE $\mathbb S_A$.add($S_A$)
            \STATE \textbf{yeild} $S_A$
            \ENDIF
            \ELSE
            \FOR{each edge expansion $r$ for $e$}
            \IF{$r$ is in the form of $e_A \to \epsilon$}
            \STATE $S_A$ $=$ \texttt{process\_synthesized}($A \to \epsilon$, $I_A$, \textbf{null}, \textbf{null})
            \IF{$S_A$ is not \textbf{null}}
            \STATE $\mathbb S_A$.add($S_A$)
            \STATE \textbf{yeild} $S_A$
            \ENDIF
            \ENDIF
            \IF{$r$ is in the form of $e_A \to e_B$}
            \STATE\COMMENT{denote the symbol of $e_B$ as $B$}
            \STATE $I_B$ $=$ \texttt{process\_left\_inherited}($A \to B$, $I_A$)
            \IF{$I_B$ is not \textbf{null}}
            \STATE $\mathbb S_B = $\texttt{CheckAttr}$(e_B, I_B, \mathbb M)$
            \FOR{each $S_B$ $\in$ $\mathbb S_B$}
            \STATE $S_A$ $=$ \texttt{process\_synthesized}($A \to B$, $I_A$, $S_B$, \textbf{null})
            \IF{$S_A$ is not \textbf{null}}
            \STATE $\mathbb S_A$.add($S_A$)
            \STATE \textbf{yeild} $S_A$
            \ENDIF
            \ENDFOR
            \ENDIF
            \ENDIF
            \IF{$r$ is in the form of $e_A \to e_B e_C$}
            \STATE\COMMENT{denote the symbol of $e_B$ as $B$ and the symbol of $e_C$ as $C$}
            \STATE $I_B$ $=$ \texttt{process\_left\_inherited}($A \to BC$, $I_A$)
            \IF{$I_B$ is not \textbf{null}}
            \STATE $\mathbb S_B = $\texttt{CheckAttr}$(e_B, I_B, \mathbb M)$
            \FOR{each $S_B$ $\in$ $\mathbb S_B$}
            \STATE $I_C$ $=$ \texttt{process\_right\_inherited}($A \to BC$, $I_A$, $S_B$)
            \IF{$I_C$ is not \textbf{null}}
            \STATE $\mathbb S_C = $\texttt{CheckAttr}$(e_C, I_C, \mathbb M)$
            \FOR{each $S_C$ $\in$ $\mathbb S_C$}
            \STATE $S_A$ $=$ \texttt{process\_synthesized}($A \to BC$, $I_A$, $S_B$, $S_C$)
            \IF{$S_A$ is not \textbf{null}}
            \STATE $\mathbb S_A$.add($S_A$)
            \STATE \textbf{yeild} $S_A$
            \ENDIF
            \ENDFOR
            \ENDIF
            \ENDFOR
            \ENDIF
            \ENDIF
            \ENDFOR
            \ENDIF
            \ENDPROCEDURE
        \end{algorithmic}
    \end{framedsized}
    \caption{The \texttt{CheckAttr} algorithm.}
    \label{fig:alg-checkattr}
\end{figure*}

\figurename~\ref{fig:alg-checkattr} shows the \texttt{CheckAttr} algorithm.
The procedure \texttt{CheckAttr} performs attribute checking on the CFL reachability graph.
The procedure \texttt{CheckAttr} checks an edge $e_A$ with an inherited attribute value $I_A$ and returns a set of synthesized attribute values $\mathbb S_A$.

To avoid repeated computation, \texttt{CheckAttr} utilizes a memoization map $\mathbb M$ to store the results of previous computations.
If the result of the current computation is already in $\mathbb M$, \texttt{CheckAttr} returns the result directly.

The procedure \texttt{CheckAttr} is to utilize a generator-like mechanism to return the synthesized attribute values.
Instead of returning all the synthesized attribute values at once, \texttt{CheckAttr} returns an iterator that yields one synthesized attribute value at a time.
This allows \texttt{CheckAttr} to return the synthesized attribute values one by one, which is more efficient than returning all the synthesized attribute values at once.

For terminal edges, \texttt{CheckAttr} invokes \texttt{process\_terminal} to compute the synthesized attribute value, where \texttt{process\_terminal} is a user-defined function for computing the synthesized attribute value of a terminal edge. If \texttt{process\_terminal} fails, \texttt{process\_terminal} returns \textbf{null}, indicating the inherited attribute of the terminal does not conform to semantic constraints, so that \texttt{CheckAttr} skips the current value and continues to check the next attribute value.

For non-terminal edges, \texttt{CheckAttr} iterates over all the possible productions and invokes \texttt{process\_left\_inherited}, \texttt{process\_right\_inherited}, and \texttt{process\_synthesized} to compute the inherited attribute value, the right inherited attribute value, and the synthesized attribute value, respectively. Here, \texttt{process\_left\_inherited}, \texttt{process\_right\_inherited}, and \texttt{process\_synthesized} are user-defined functions for computing the inherited attribute value, the right inherited attribute value, and the synthesized attribute value of a non-terminal edge, respectively. Similarly, these functions return \textbf{null} if the attribute value does not conform to semantic constraints, so that \texttt{CheckAttr} skips the current value and continues to check the next attribute value.

The optimizations of \texttt{CheckAttr} are implemented as follows.

\textbf{Pruning.} \texttt{CheckAttr} only continues to check the attribute if the inherited attribute value conforms to semantic constraints. Otherwise, \texttt{CheckAttr} skips the current value and continues to check the next attribute value. This is implemented by returning \textbf{null} in \texttt{process\_terminal}, \texttt{process\_left\_inherited}, \texttt{process\_right\_inherited}, and \texttt{process\_synthesized} when the attribute value does not conform to semantic constraints.

\textbf{Merging.} \texttt{CheckAttr} stores a list of synthesized attributes in $\mathbb S_A$. The list $\mathbb S_A$ will remove duplicated synthesized attributes before returning. This is implemented by using an extra set to store the synthesized attributes in $\mathbb S_A$.

\textbf{Memoization.} \texttt{CheckAttr} stores the results of previous computations in $\mathbb M$. If the result of the current computation is already in $\mathbb M$, \texttt{CheckAttr} returns the result directly.

With the help of \texttt{CheckAttr}, \techname can perform attribute checking on the CFL reachability graph and find the shortest path that satisfies all the attribute constraints.